\documentclass[a4paper,11pt]{article}
\usepackage{pos}

\def\XXint#1#2#3{{\setbox0=\hbox{$#1{#2#3}{\int}$ }
\vcenter{\hbox{$#2#3$ }}\kern-.6\wd0}}

\usepackage{slashed}

\usepackage{multirow}

\usepackage{float}

\usepackage{stackrel}

\title{Opportunities at FCC-ee for quark \& lepton flavour physics}
\ShortTitle{Flavour at FCC-ee}

\author*[a]{Luiz~Vale~Silva}

\affiliation[a]{\it Departamento de Matem\'{a}ticas, F\'{i}sica y Ciencias Tecnol\'{o}gicas,

Universidad Cardenal Herrera-CEU, CEU Universities,

46115 Alfara del Patriarca, Val\`{e}ncia, Spain}

\emailAdd{luiz.valesilva@uchceu.es}

\abstract{
The FCC-ee phase of a Future Circular Collider is generating great interest due to its versatility, allowing the study of various electroweak thresholds, $Z$, $WW$, $ZH$, and $t \bar{t}$. Electroweak precision physics is complemented by flavour physics measurements based on the unprecedented statistics attainable at the $Z$ pole, and benefiting from the low-background experimental environment (similar to Belle~II), and from the production of the full spectrum of hadron species together with large boosts (similar to LHCb). A wide range of measurements is possible, spanning a rich variety of physics cases in both quark and lepton flavour physics sectors. Other electroweak thresholds can also be considered in this endeavour. A commensurate effort from the theory community will be needed to interpret future measurements. I present an overview of the broad potential of the FCC-ee flavour physics program.
}

\FullConference{The European Physical Society Conference on High Energy Physics (EPS-HEP2025)\\
7-11 July 2025\\
Marseille, France\\}

\begin{document}

\maketitle

Flavour physics played a crucial role in building the Standard Model (SM) of particle physics.
Moreover, the study of quark and lepton processes permits further testing fundamental properties of the SM such as: the structure of couplings of flavour-changing charged currents, the absence of Flavour-Changing Neutral Currents (FCNCs) at the tree level and its loop-generated dynamics, the universality of lepton couplings across generations, and the huge suppression of charged-Lepton Flavour Violation (cLFV),
to give examples aligned with these notes.
In testing the SM, one can
focus on the comparison of precise experimental measurements and theoretical predictions, which then sets constraints on possible NP contamination;
one can also consider cases in which the SM prediction is very much below
the experimental sensitivity, and in which there remains a lot of room for NP contributions; etc.

There is a growing effort to evaluate physics cases for FCC-ee, the electron-positron phase of the Future Circular Collider \cite{FCC:2018byv,FCC:2018evy,FCC:2025lpp}, thus strengthening the motivation to build it.
In this document we focus on the scope of flavour applications.
FCC-ee has a huge potential for flavour physics, since in particular it can produce all sorts of heavy flavours ($B$ mesons, $B_s$ mesons, $B_c$ mesons, $b$-baryons, apart from all sorts of charmed hadrons and $\tau$ leptons) from the production and decay of on-shell $\sim 6 \times 10^{12}$ $Z$ bosons.\footnote{The full list of particles which trigger the interest of particle physicists also includes exotic states, whose production remains to be estimated in the context of FCC-ee. It also remains to be seen its potential impact on the study of strange physics, e.g., $K_S \to \mu^+ \mu^-$ and hyperons. See Ref.~\cite{Ai:2024nmn} for discussions in the context of CEPC.}
As we will see, there are also flavour physics cases associated with $W$, $H$, and top-quark production and decay.
FCC-ee accumulates many other interesting properties for the study of different flavour processes, such as high boost, negligible trigger losses, high geometrical acceptance, low backgrounds, flavour-tagging power, initial energy constraint, and production of polarized baryons \cite{Monteil:2021ith,Grossman:2021xfq,Novotny:2022qgg,Cobal:2022jep,Zuo:2024grn,deBlas:2024bmz}.
In the following we review some applications of FCC-ee given its evolving specifications, and point out its qualities according to different processes of physical interest.

We start with studies of the CKM matrix.
In the quark sector there is a single phase that must account for CP violating phenomena in different flavour sectors and categories of observables.
The present overall consistent picture \cite{UTfit:2022hsi,ValeSilva:2024jml} has only been possible thanks to very precise measurements and theoretical predictions. In particular, one must have a good control over the QCD dynamics, which is ubiquitous when dealing with quark physics.
Increasing the experimental precision will require a commensurate effort from the theory community, e.g., in addressing QCD, and also QED, effects.

Of foremost importance, the redundancy of observables offers complementary tests of the CKM matrix in the SM.
One has to consider FCC-ee in the context of expected improvements from LHCb upgrades and Belle II.
The extraction of some angles of the unitarity triangle is expected to improve by a modest but important factor of $2$ compared to projections from LHCb 300/fb and Belle II 50/ab \cite{ATLAS:2025lrr}.
With increasing experimental precision comes the need to worry about sub-leading theoretical uncertainties. This is the case of penguin pollution in the extraction of $\beta$ $(\phi_1)$ from the so-called golden mode $B \to J/\psi K_S$,
which is in fact already a pressing issue.
Furthermore,
the extraction of the angle $\alpha$ $(\phi_2)$ is based on an isospin analysis relating different isospin modes of $B$s to $\pi$s and $\rho$s; FCC-ee can address in particular the mode with neutral pions, and by looking at modes with $\eta$s and $\eta'$s it may be possible to have a better control over isospin breaking corrections \cite{Wang:2022nrm}.
Other angles are also expected to have an increased experimental precision, such as $\beta_s$ and else \cite{Aleksan:2021gii,Aleksan:2021fbx,Li:2022tlo,Aleksan:2024jwm,Peng:2025bki}.

Shifting to CKM magnitudes, from the $\sim 3 \times 10^{8}$ $W$-pair dataset one can envisage extractions of $|V_{cb}|$ and $|V_{cs}|$ \cite{Marzocca:2024mkc,Liang:2024hox}. This relies on a good tagging efficiency of the boosted quark flavours. With tagging efficiency uncertainties as low as $1\%$ the uncertainty in the extraction of $|V_{cb}|$ could be as low as $0.5\%$ \cite{Marzocca:2024mkc}.
Although more studies will assess further expected uncertainties attached to this extraction,
these are certainly good news, since this measurement does not rely on the usual semi-leptonic $B$ decay modes, that are presently difficult to control with such a low uncertainty, and in any case the extraction of $|V_{cb}|$ based on $W \to b c$ consists of an orthogonal measurement. With a tagging efficiency uncertainty of about $1\%$ a sub-percent uncertainty is also expected in the case of $|V_{cs}|$ \cite{Marzocca:2024mkc}, which is comparable to projections from the STCF based on leptonic decays \cite{Liu:2021qio}. Let us also point out that the analysis of $B_c \to \tau \nu_\tau$ does not offer a precise extraction of $|V_{cb}|$ due to uncertainties in the hadronization fraction;\footnote{The use of the semi-leptonic decay $B_c \to J/\psi \ell \nu_\ell$ as a normalization mode would cancel the dependence on $|V_{cb}|$. On the other hand, the branching ratio of the mode $B_c \to D^{(\ast)} \ell \nu_\ell$
scales with $|V_{ub}|^2$, and so has a much lower statistics.}
conversely, given $|V_{cb}|$, the $B_c$ leptonic decay mode could be used in the determination of this hadronic quantity.
Using a similar strategy, no improvement is expected for $|V_{ub}|$, which is more than $10$ smaller than $|V_{cb}|$.
The extraction of $|V_{ts}|$ is also on the horizon \cite{Xunwu}.

\begin{figure}
    \centering
    \includegraphics[scale=0.26]{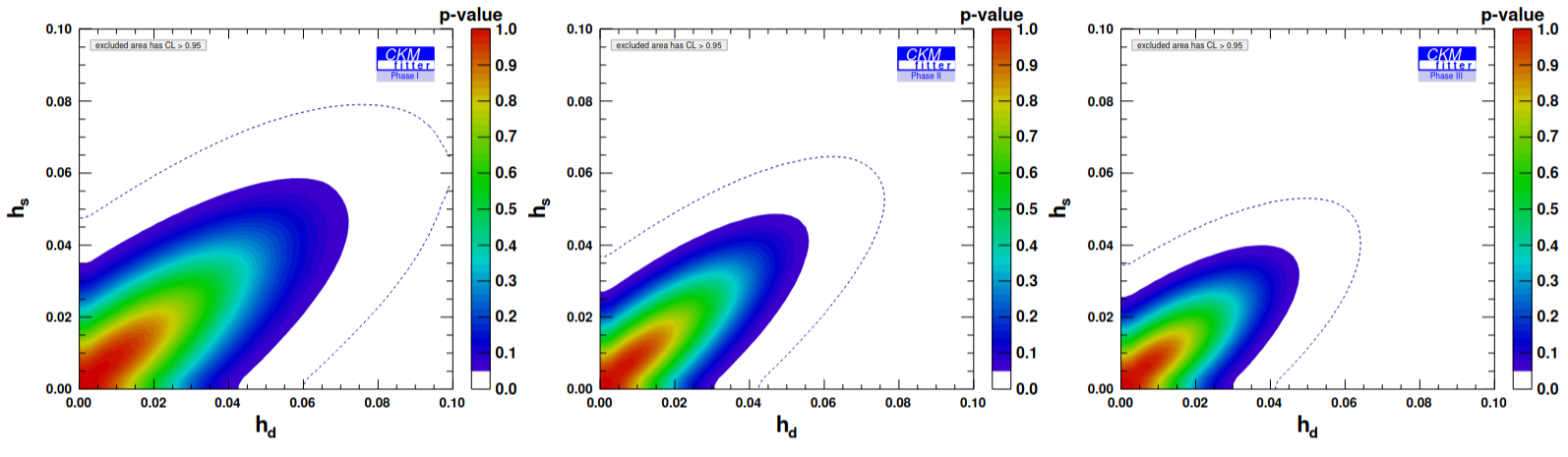}
    \caption{
    Bounds on the relative sizes of NP with respect to the SM (for which $h_d=h_s=0$) in the $B_d$ and $B_s$ meson systems. Current bounds reach $\sim 20\%$.
    Phase I ($\sim$ 2030’s) corresponds to Belle II 10/ab \& LHCb 50/fb; Phase II ($\sim$ 2040’s) to Belle II 50/ab \& LHCb 300/fb; and Phase III ($\sim$ 2050’s) corresponds to Phase II, together with improvements from FCC-ee.
    The different observables used in these fits are considered to perfectly agree.
    }
    \label{fig:fig}
\end{figure}

In global CKM matrix fits, processes at the quantum level play a major role. Such processes often set very strong constraints on NP, pushing its characteristic scale to very high energies. To give an example, consider NP in processes that change flavour by two units, which in the SM are mediated by the usual box diagrams. We parameterize NP contributions with two parameters per neutral meson system: $h$ gives the relative size of NP compared to the SM, while $\sigma$ allows for new CP-violating phases. We consider that NP is heavy with respect to the mass scale of the neutral meson, and thus one possible way of introducing NP is through $|\Delta F|=2$ effective contact interactions of dimension-6.
Since so many precision observables are available, one can consider extracting the Wolfenstein parameters of the CKM matrix altogether with the allowed level of NP contamination.
As of now, one observes a substantial degrading in precision in the extraction of the Wolfenstein parameters, and a sizable room for NP \cite{Charles:2020dfl}.
As seen from Fig.~\ref{fig:fig}, the constraints on the relative sizes of NP steadily improve when moving into future projections.
To achieve progress beyond these expectations,
two key sets of quantities are QCD inputs and $|V_{cb}|$.
It is worth mentioning that
although semi-leptonic asymmetries do not play an important role in this study, they are expected to improve substantially with FCC-ee \cite{Stephane,Charles:2020dfl}, which would have an interesting effect in constraining well-motivated NP extensions \cite{Miro:2024fid}.
Improvements in the kaon sector also depend on QCD inputs and the value of $|V_{cb}|$.

Starting with bottom physics, let us now discuss decays mediated by FCNCs, that are rare in the SM due to both GIM and CKM suppressions.
Since the bottom quark belongs to the third generation, it is generally expected that this sector is more sensitive to NP \cite{Allwicher:2025bub}.
A key flavour physics case for FCC-ee is $b \to s \tau \tau$ \cite{Kamenik:2017ghi}, that thus also involves leptons of the third generation. There are currently orders of magnitude separating SM predictions and experimental measurements. With an excellent vertexing system, the kinematics of the $B$ decay can be reconstructed, and the sensitivity goes close to performing measurements of SM rates \cite{Li:2020bvr,Miralles}. With a statistics higher than currently projected, one can perform the determination of observables beyond the total branching fraction, such as angular observables and observables based on the reconstructed polarizations of the $\tau$s \cite{Kamenik:2017ghi}.

A general class of measurements that FCC-ee can improve consists of processes carrying invisible particles in the final states, such as the neutrinos from the decays of $\tau$ leptons in the last paragraph.
There are plenty of other interesting $b$-decay modes that FCC-ee can analyze. This is the case for modes mediated by contact interactions having neutrinos such as those carrying the flavours $ (b s) (\nu_1 \nu_2) $ \cite{Amhis:2023mpj,AlvarezCartelle:2025mtx},
which are relevant for model building since left-handed neutrinos can be related to charged leptons via weak-isospin symmetry. The precision in branching ratios is expected to be substantially better than Belle II can achieve.
In the case of light charged leptons in the final state, FCC-ee can improve in particular the electron mode,
which is crucial for further improving tests of lepton universality \cite{Bordone:2025cde}.
With high statistics one can study CP violation based on time-dependent observables \cite{Kwok:2025fza}.
Bottom baryon decays can also be used to test the SM and look for the presence of NP \cite{Altmannshofer:2025eor,Beck:2025bgc}.

Apart from rare decay modes, FCC-ee can look at $B_{(c)} \to \tau \nu_\tau$ transitions. Here as well invisible particles are present, and FCC-ee can perform measurements based on the reconstruction of the full event, using information from the companion bottom quark and the vertexing system \cite{Amhis:2021cfy,Zuo:2023dzn}. This is, e.g., important in view of the long-standing LFU tension in the $b \to c$ category of decays \cite{Ho:2022ipo}.

Let us now quickly move to charm physics.
The number of produced charm quarks is comparable to bottom quarks, which means a large statistical sample also applies.
We point out two categories of observables, namely, direct CP violation and rare charm-meson decays.
Direct CP violation was discovered by LHCb in $ \Delta A_{CP} = A_{CP} (D^0 \to K^- K^+) - A_{CP} (D^0 \to \pi^- \pi^+) $ \cite{LHCb:2019hro} (see also Ref.~\cite{LHCb:2022lry}), which will further improve substantially the uncertainty attached to $\Delta A_{CP}$ in the future \cite{ATLAS:2025lrr}.
However, it is very difficult to predict this observable in the SM due to non-perturbative QCD effects \cite{Khodjamirian:2017zdu,Pich:2023kim}.
It is then fundamental to perform measurements of related modes to draw a clearer picture of the underlying mechanism producing CP violation in branching ratio asymmetries.
Since its production asymmetry is not at the same level of a proton-proton collider,
FCC-ee could provide an important additional measurement for such a suppressed quantity as $ \Delta A_{CP} $.
Moreover, it can measure neutral $\pi$s in the final state, which are related to the measurement by LHCb via isospin.
There is an ongoing effort to establish FCC-ee capabilities \cite{Dominik_CKM}; for instance, an $\mathcal{O} (10^{-2} \%)$ sensitivity to the CP asymmetry in $D^0 \to \pi^0 \pi^0$ is currently estimated.
Other related modes, such as multi-body and $D \to \gamma \gamma$ decays are also of theoretical interest \cite{Jernej,Tommaso}.
In the case of rare decays, similarly to the bottom sector, FCC-ee can address electron and neutrino modes. It can also measure $\pi^0$s in the final state, which would help better understanding the non-perturbative dynamics underlying processes with two charged pions in the final state \cite{Fajfer:2023tkp}, for which LHCb has provided plenty of data in the muonic decay mode \cite{LHCb:2021yxk}. The set of physics cases also includes di-neutrino modes. To give a concrete example, the bound on the mode $ D^0 \to \pi^0 \nu \bar\nu $ \cite{BESIII:2021slf} is expected to improve by as much as a factor of $100$ based on a naive rescaling of the luminosity \cite{Bause:2020xzj}. A very recent analysis of $ D \to \pi \pi \nu \bar\nu $ establishes a sensitivity to the branching ratio of $\mathcal{O} (10^{-7})$ \cite{DiCanto:2025fpk}.
There are also interesting physics cases for
baryon decay modes, such as $\Lambda_c \to p \ell^+ \ell^-$, $\ell = e, \mu$, see Ref.~\cite{DiCanto:2025fpk}.

Moving to tau physics, large yields are again expected. As illustrated by LEP, which provides measurements that are still relevant today,
an electron-positron machine can provide outstanding measurements based on highly boosted taus,
due to a better control of the hadronic background, a better use of $\tau$-pair production, and a better control of the kinematics \cite{Dam:2018rfz,Pich:2020qna,Lusiani}.
From the theoretical side, as previously mentioned the $\tau$ is interesting for the study of third generation lepton dynamics, but also as a laboratory for QCD, via the study of inclusive and exclusive branching ratios \cite{Pich:2020qna}.
Moreover, FCC-ee can improve the determination of $\tau$ properties such as its mass, lifetime, and branching ratios. This has a huge impact on the study of lepton universality tests such as $g_\tau = g_\mu$, as illustrated by the expression

\begin{equation}
    \left( \frac{g_\tau}{g_\mu} \right)^2 \propto \frac{\mathcal{B} (\tau \to e \nu \bar\nu)}{\mathcal{B} (\mu \to e \nu \bar\nu)} \cdot \frac{\tau_\mu m_\mu^5}{\tau_\tau m_\tau^5} \,,
\end{equation}
i.e., to study universality in lepton couplings one needs to know very well not only the $\tau \to e \nu \bar\nu$ branching ratio, but also the mass $m_\tau$ and lifetime $\tau_\tau$.

Finally, plenty of tests of cLFV will be pushed to higher sensitivity. These processes are virtually absent in the SM, and thus they set essential tests of its leptonic sector. In the case of $\tau$ decays, one can benefit from the absence of neutrinos in the final state in some possible decay modes to reconstruct candidate events. A substantially higher sensitivity is expected compared to other machines.
There are various other possible tests in $Z$, Higgs and heavy quark decays, see, e.g., Refs.~\cite{Qin:2017aju,Calibbi:2021pyh}. In the case of $Z \to \mu e$, a crucial aspect setting the sensitivity is the control of the background stemming from muons being misidentified as electrons \cite{Dam:2018rfz}.
In the case of bottom quark decays, an easier reconstruction is expected due to a smaller number of invisible particles in the final state, such as in the case of $b \to s \tau \ell$, $\ell = e, \mu$, if compared to the di-tau decay mode \cite{Chrzaszcz:2021nuk}.
The decay $D \to \tau e$ could also be investigated.
Flavour changing neutral Higgs decays to $ b s $ or $ c u $ can generally be probed better with $e^+ e^- \to Z H$ than via indirect searches \cite{Kamenik:2023hvi,Arroyo-Urena:2025mju}.
High-$p_T$ searches for $e^+ e^-$ into different quark and lepton flavours can also be considered in testing FCNCs \cite{Shi:2019epw,Altmannshofer:2023tsa,Greljo:2024ytg}.
The reach of tests of baryon number violation such as in $\tau$ decays remains to be estimated.

\vspace{3mm}

In conclusion, we attempted to convince the reader that FCC-ee has a lot to offer to flavour physics, both quark and lepton sectors. The specific projections depend on the detecting properties, and flavour physics sets requirements that must be taken into account. In both sectors, very important tests of SM aspects can be performed, and especially in the case of quark flavour physics (but also tau physics) in order to match the experimental accuracy theoretical progress will be needed, part of which will be experimentally driven.
The previous subset of physics cases is by no means exhaustive. Further work is needed in order to draw a more comprehensive picture of FCC-ee capabilities in addressing flavour physics observables.

\vspace{3mm}

\noindent
\textbf{Acknowledgements.}
First and foremost, I thank the FCC collaboration for the opportunity to present this review.
Also, special thanks to
St\'{e}phane Monteil, Dominik Suelmann and Xunwu Zuo for many clarifying exchanges.
Finally, I thank
Marcella Bona, Claudia Cornella, Timothy Gershon, Vladimir Gligorov, Mart\'{i}n Novoa-Brunet, Chris Parkes, and Ludovico Vittorio for the lively discussion following the talk at the origin of these proceedings.
This work is supported by the Spanish Government (Agencia Estatal de Investigaci\'{o}n MCIN/AEI/10.13039/501100011033) Grants No. PID2020–114473GB-I00 and No. PID2023-146220NB-I00, and CEX2023-001292-S (Agencia Estatal de Investigaci\'{o}n MCIU/AEI (Spain) under grant IFIC Centro de Excelencia Severo Ochoa).

\bibliography{mybib}{}
\bibliographystyle{unsrturl}

\end{document}